\documentclass[12pt,preprint]{aastex}

\slugcomment{Accepted  for publication in ApJ}

\begin{document}

\title{Observation of TeV gamma-rays from the unidentified source HESS J1841-055 with the ARGO-YBJ experiment}


\author{B.~Bartoli\altaffilmark{1,2},
 P.~Bernardini\altaffilmark{3,4},
 X.J.~Bi\altaffilmark{5},
 I.~Bolognino\altaffilmark{6,7},
 P.~Branchini\altaffilmark{8},
 A.~Budano\altaffilmark{8},
 A.K.~Calabrese Melcarne\altaffilmark{9},
 P.~Camarri\altaffilmark{10,11},
 Z.~Cao\altaffilmark{5},
 R.~Cardarelli\altaffilmark{11}
% ...
\\ (The ARGO-YBJ Collaboration)}


 \altaffiltext{1}{Dipartimento di Fisica dell'Universit\`a di Napoli
                  ``Federico II'', Complesso Universitario di Monte
                  Sant'Angelo, via Cinthia, 80126 Napoli, Italy.}
 \altaffiltext{2}{Istituto Nazionale di Fisica Nucleare, Sezione di
                  Napoli, Complesso Universitario di Monte
                  Sant'Angelo, via Cinthia, 80126 Napoli, Italy.}
 \altaffiltext{3}{Dipartimento Matematica e Fisica "Ennio De Giorgi",
                  Universit\`a del Salento,
                  via per Arnesano, 73100 Lecce, Italy.}
 \altaffiltext{4}{Istituto Nazionale di Fisica Nucleare, Sezione di
                  Lecce, via per Arnesano, 73100 Lecce, Italy.}
 \altaffiltext{5}{Key Laboratory of Particle Astrophysics, Institute
                  of High Energy Physics, Chinese Academy of Sciences,
                  P.O. Box 918, 100049 Beijing, P.R. China.}
 \altaffiltext{6}{Dipartimento di Fisica dell'Universit\`a di
                  Pavia, via Bassi 6, 27100 Pavia, Italy.}
 \altaffiltext{7}{Istituto Nazionale di Fisica Nucleare, Sezione di Pavia,
                  via Bassi 6, 27100 Pavia, Italy.}
 \altaffiltext{8}{Istituto Nazionale di Fisica Nucleare, Sezione di
                  Roma Tre, via della Vasca Navale 84, 00146 Roma, Italy.}
 \altaffiltext{9}{Istituto Nazionale di Fisica Nucleare - CNAF, Viale
                  Berti-Pichat 6/2, 40127 Bologna, Italy.}
 \altaffiltext{10}{Dipartimento di Fisica dell'Universit\`a di Roma ``Tor Vergata'',
                   via della Ricerca Scientifica 1, 00133 Roma, Italy.}
 \altaffiltext{11}{Istituto Nazionale di Fisica Nucleare, Sezione di
                   Roma Tor Vergata, via della Ricerca Scientifica 1,
                   00133 Roma, Italy.}
% ...

\begin{abstract}

We report the observation of a very high energy $\gamma$-ray source, whose position is coincident with HESS J1841-055.
This source has been observed for 4.5 years by the ARGO-YBJ experiment from November 2007 to July 2012. Its emission is detected with a statistical significance of 5.3 standard deviations. Parameterizing the source shape with a two-dimensional Gaussian function we estimate an extension
$\sigma=(0.40^{+0.32}_{-0.22})^{\circ}$, consistent with the HESS measurement.
The observed
energy spectrum is $dN/dE =(9.0\pm1.6) \times 10^{-13}(E/5 TeV)^{-2.32\pm0.23}$ photons cm$^{-2}$ s$^{-1}$ TeV$^{-1}$, in the energy range
0.9-50 TeV. The integral $\gamma$-ray flux above 1 TeV is 1.3$\pm$0.4 Crab units, which is 3.2$\pm$1.0 times the flux derived by HESS. The differences in the flux determination between HESS and ARGO-YBJ, and
possible counterparts at other wavelengths are discussed.

\end{abstract}


\section{Introduction}
Very High Energy (VHE) $\gamma$-ray astronomy opened a new window to explore the extreme non-thermal phenomena in the universe.
VHE $\gamma$-rays are tracers of non-thermal particle acceleration and are used to probe the conditions and the underlying
astrophysical processes inside their sources. In the past decade, great progresses have been made in the field of VHE $\gamma$-ray astronomy.
More than one hundred VHE $\gamma$-ray  emitters have been detected, belonging to several categories: Active
Galactic Nuclei (AGNs), Pulsar Wind Nebulae (PWNs), SuperNova Remnants (SNRs),  X-ray Binaries
(XBs), and starburst galaxies. However, there is a fraction of VHE sources still unidentified because do not appear to have obvious counterparts at other wavelengths. This kind of sources may constitute a new class of objects with
different emission properties.

HESS J1841$-$055 is an enigmatic unidentified VHE $\gamma$-ray source discovered by the HESS collaboration during the Galactic plane survey \citep{aharon08}. Its image shows a high extension, the measured axes for an elongated two-dimensional  Gaussian shape being
$0.41^{\circ}\pm0.04^{\circ}$ (major) and
$0.25^{\circ}\pm0.02^{\circ}$ (minor).
HESS J1841$-$055, therefore, is one of the most extended sources in the VHE $\gamma$-ray band.
The spectrum is best fitted by a simple power law with photon index $\alpha=-2.41\pm0.08$ in the energy range from 0.54 TeV to 80 TeV. The integral flux is
$9.1\times 10^{-12}$ photons cm$^{-2}$ s$^{-1}$ at energies above 1 TeV, about 40.3\% of the Crab unit \citep{aharon06}.

To date, no obvious counterpart has been found at other wavelengths. The wide
VHE $\gamma$-ray morphology suggests
that HESS J1841$-$055 may be the blend of multiple sources. \cite{aharon08}
found four candidates which could be responsible for at least part of the entire VHE $\gamma$-ray
emission: the two pulsars PSR J1841$-$0524 and PSR J1838$-$0549, the diffuse source G26.6$-$0.1, which is a candidate SNR basing on its ASCA spectrum,
and finally the high-mass XB AX J1841.0$-$0536. Basing on a striking spatial correlation, \cite{sguer09}  propose that the
Supergiant Fast X-ray Transient (SFXT) AX J1841.0$-$0536  could be responsible for at least a fraction of the VHE $\gamma$-ray emission from HESS
J1841$-$055, thus being the prototype of a new class of Galactic transient
MeV/TeV emitters.
Using $\gamma$-rays with energies $>100$ GeV detected by $Fermi$-LAT,
\cite{neron10} found an event cluster adjacent to HESS J1841$-$055, and a more extended event cluster at the opposite side. This may be an evidence that HESS J1841$-$055 is composed of at least two different components.
On the other hand, \cite{neron12} suggest an association only with PSR J1841$-$0524, which is situated in the center of the extended source.

The ARGO-YBJ experiment is an air shower array with large
field of view (FOV) which continuously monitor the northern sky. The emission from the Crab Nebula has been detected with a statistical
significance of 17 standard deviations (s.d.) at energies around 1 TeV. With such a sensitivity, other 4 known VHE $\gamma$-ray sources have been detected with significance greater than 5 $\sigma$: Mrk 421 \citep{barto11}, Mrk 501 \citep{barto12a}, and the two extended sources MGRO J2031+41 \citep{barto12b} and MGRO J1908+06 \citep{barto12c}. It should be pointed out that the fluxes of the two extended sources measured by the EAS arrays Milagro and ARGO-YBJ are much higher than that determined by the Cherenkov arrays, showing that there are some systematic differences between the two observation techniques for extended sources \citep{barto12b,barto12c,abdo12}. Since also HESS J1841$-$055 is an extended source, its study would benefit from an observation using EAS arrays. HESS J1841$-$055 is observed by ARGO-YBJ, at the edge of its FOV, 4.8 hours per day with zenith angle less than 50$^{\circ}$, culminating at 35.7$^{\circ}$.
This work presents the observation results for HESS J1841$-$055 with
the ARGO-YBJ experiment.

\section{The ARGO-YBJ experiment}
The ARGO-YBJ experiment is a full coverage
extensive air shower array resulting from a collaboration
between Chinese and Italian institutions and is  designed for VHE $\gamma$-ray astronomy and cosmic ray observations.
The detector is operating at the Yangbajing International Cosmic Ray
Observatory  (Tibet, P.R. China), at an altitude of 4300 m a.s.l..
The detector, extensively described
in \citep{aielli06,aielli09c}, consists of a single layer of Resistive Plate Chambers (RPCs, 2.8 m $\times$1.25 m), equipped with 10 logical pixels (called ``pads'', 55.6 cm $\times$ 61.8 cm) used for triggering and timing purposes. 130 clusters (each composed by 12 RPCs) are installed to form the central carpet of 74 m $\times$ 78 m with an active area of $\sim$93\%, surrounded by 23 additional clusters (``guard ring''). The total area of the array is 110 m $\times$ 100 m.  The arrival
time of the particles is measured by time-to-digital converters (TDCs) with a resolution of about
1.8 ns \citep{aielli09c}. To calibrate the 18,360 TDC
channels,  a software  method has been developed using cosmic ray showers \citep{he07}.
The calibration precision is 0.4 ns  and the procedure is applied every month \citep{aielli09}.

The central 130 clusters started taking data in July 2006,
while the complete ARGO-YBJ detector including the ``guard ring'' collected
data since November 2007. The RPC carpet is connected to two independent data acquisition
systems, corresponding to the shower and scaler operation modes \citep{aielli08}.
In the current work, only data from the shower mode are used.
In shower mode, the ARGO-YBJ detector is operated by requiring at least 20 fired pads
($N_{pad}$) within 420 ns on the entire carpet detector.
The trigger rate is 3.5 kHz with a dead time of 4\%
and the average duty-cycle is higher than $86\%$.

The high granularity of the apparatus allows a complete and detailed space-time three-dimensional  reconstruction of the
shower profile and therefore of the incident direction of the primary particle.
Through the analysis of the position, size and shape of the reconstructed Moon
 and Sun shadows in the cosmic ray flux, the angular resolution, pointing accuracy and stability of
the ARGO-YBJ detector array have been thoroughly tested  \citep{barto11b,aielli11}.
The Point Spread Function (PSF) is quantified using a parameter $\psi_{70}$ as
the opening angle containing 71.5\% of the events.
For cosmic ray-induced air showers $\psi_{70}$ is 2.8$^{\circ}$ for
$N_{pad}\sim 20$, while becomes 0.47$^{\circ}$ for
$N_{pad}>1000$ \citep{barto11,barto11b}, in good agreement with Monte Carlo
predictions. The simulations show that the angular resolution for
$\gamma$-induced showers is 30$-$40\% smaller. The effective
area of the detector for $\gamma$-induced showers depends on the
$\gamma$-ray energy and incident zenith angle, e.g., it is about 100 m$^2$
at 100 GeV and $>$10,000 m$^2$ above 1 TeV for a zenith angle of
20$^{\circ}$ \citep{aielli09b}.

\section{Data analysis}
The data set used in this analysis refers to
the period from November 2007  to July 2012. The total effective observation time is 1492.6 days.
For the analysis presented
in this paper, only events with a zenith angle less than 50$^{\circ}$ are used, and the data set is
divided into six groups according to $N_{pad}$. To achieve a good angular resolution, the event selections used in \citep{barto11} are applied here.
The total number of events after filtering used in this
work is $2.42\times10^{11}$.
The opening angles $\psi_{70}$ for events with $N_{pad}>60$ and $N_{pad}>100$ are 1.36$^{\circ}$ and 0.98$^{\circ}$,
respectively.
For the data set in each group,
 the whole sky map in celestial coordinates (right ascension and declination) is divided into a grid of
$0.1^{\circ}\times0.1^{\circ}$ bins and filled with detected events according to their
reconstructed arrival direction.
The ``direct integral method'' \citep{fleysher04} is adopted to estimate the cosmic-ray background and
to extract the excess of $\gamma$-induced
showers from each bin.
The correction procedure described in \citep{barto11} has been applied
to remove the effect of cosmic ray anisotropy on a scale of $11^{\circ}\times11^{\circ}$. A Gaussian smoothing method is used to take into
account the PSF of the ARGO-YBJ detector. That is, the events in a circular area centered on the bin
with an angular radius of 1.3$\psi_{70}$, are summed after weighting
with the Gaussian-shaped PSF.
The Li-Ma method \citep{li83} is used to estimate the
significance of the excess in each bin.

With this procedure, the northern sky has been surveyed \citep{chensz11}. The significance of the excess observed from the direction of  the Crab Nebula is 17 s.d., indicating that the cumulative 5 s.d. sensitivity of ARGO-YBJ has reached 0.3 Crab unit for point sources.  The sensitivity is  dependent on the declination of the source, being degraded by a factor 3.5 at the declination of HESS J1841$-$055 \citep{chensz11}. For an extended source with a symmetrical two-dimensional Gaussian shape with  $\sigma=0.40^{\circ}$, the sensitivity is degraded by 15\%. Therefore, a simple estimation indicates that the flux from HESS J1841$-$055 should be about 1.2 Crab units in order to be detected by ARGO-YBJ with 5 s.d.. The required flux slightly varies if the spectrum is different from that of the Crab Nebula.

\section{Results}
The significance map around HESS J1841$-$055, as observed by ARGO-YBJ using events with $N_{pad}>60$, is shown in Figure 1. For
comparison, the twelve sources in the second $Fermi$-LAT catalog \citep{abdo12a} around HESS J1841$-$055 are also
marked in the figure. Weak excesses are observed along the Galactic plane,
indicating a diffuse $\gamma$-ray emission.
An analysis of the diffuse $\gamma$-ray emission
using ARGO-YBJ data can be found in \citep{ma11}. The highest
significance is 5.3 s.d. at $\alpha$=$18^h39^m$ and $\delta$=-6$^{\circ}3'$ (J2000), which is displaced  0.7$^{\circ}$ from the center of HESS J1841$-$055. To estimate the statistical error of the position, the data are sampled 20,000 times and the statistical errors in both directions are about 0.45$^{\circ}$. However, most of the excesses overlap the extended region of HESS J1841$-$055 and its gravity center ($\alpha$=$18^h40^m\pm12^m$ and $\delta$=-5$^{\circ}52'\pm13'$), obtained using all the pixels with significance greater than 3 s.d. within $3^{\circ}\times3^{\circ}$ around HESSJ1841$-$055, is 0.4$^{\circ}$ off the center of HESS J1841$-$055.
These displacements may be caused by different concurring effects beside fluctuation. (1) Complex morphology. According to the HESS result, HESS J1841$-$055 possibly has two or three peaks and the positions of the two largest ones are both 0.44$^{\circ}$ off the center. (2) The systematic pointing error of ARGO-YBJ is 0.2$^{\circ}$, slightly increasing at the boundary of the ARGO-YBJ FOV. (3) The contribution of the nearby VHE source HESS J1837$-$069, partially containing its emission. Therefore, the signal position observed by ARGO-YBJ largely overlaps
HESS J1841$-$055.

The intrinsic extension of HESS J1841$-$055 is determined by fitting the distribution of $\theta^{2}$ for the events exceeding the background as shown in Figure 2, where $\theta$ is the angular distance of each event to the position of HESS J1841$-$055. To achieve a good angular resolution, only events with $N_{pad}>100$ are used in this fit. In order to fit the data, a set of $\gamma$-rays is generated taking into account the Spectral Energy Distribution (SED), the intrinsic source extension, and the detector PSF. The extension is estimated by minimizing the $\chi^2$ between data and generated events, from 0$^{\circ}$ to 1$^{\circ}$ with steps of 0.1$^{\circ}$.
Assuming a spectral index $-$2.3, the intrinsic extension is determined to be $\sigma_{ext}=(0.40_{-0.22}^{+0.32})^{\circ}$. It is found that the dependence on the SED is negligible within the uncertainties. This result is consistent with the estimation by the HESS collaboration,
i.e., $0.41^{\circ}\pm0.04^{\circ}$ and $0.25^{\circ}\pm0.02^{\circ}$ along the major and minor axes, respectively \citep{aharon08}.

Assuming an intrinsic extension $\sigma_{ext}=0.40^{\circ}$, we estimate the spectrum of HESS J1841$-$055 using the ARGO-YBJ data with the conventional fitting method described in \citep{barto11}. In this procedure, the expectation function is generated by sampling events in the energy range from 10 GeV to 100 TeV and taking into account the detailed ARGO-YBJ detector response, assuming a power law with its spectral index as a parameter. We define five $N_{pad}$ intervals: 60$-$59,100$-$199, 200$-$499, 500$-$999, and $\geq$1000.
The best fit to the SED and the corresponding 1 $\sigma$ error region are shown in Figure 3. The differential flux (TeV$^{-1}$ cm$^{-2}$ s$^{-1}$) in the energy range from 0.9 TeV to 50 TeV is
\begin{equation}
\frac{dN}{dE}=(9.0\pm1.6) \times 10^{-13}(E/5\; TeV)^{-2.32\pm0.23}.
\end{equation}
The median energies of the five $N_{pad}$ intervals are 2.3, 3.5, 7.1, 14 and
22  TeV, respectively.
The integral flux is 1.3$\pm$0.4 Crab units at energies above 1 TeV, which is 3.2$\pm$1.0 times the flux  determined by the HESS experiment, i.e.,
0.40 Crab unit.

\section{Discussion}
The integrated energy flux above 1 TeV measured by ARGO-YBJ is
$\sim 1.9\times 10^{-10}$ erg cm$^{-2}$ s$^{-1}$, corresponding to a source
luminosity, assuming isotropic emission, of
L($>$1 TeV)$\sim 2.3\times 10^{34}$(D/1 kpc)$^2$ erg s$^{-1}$, where $D$ is
the distance to the source. However, due to the limitations in the angular
resolution, this flux may also include other contributions apart from
HESS J1841$-$055.
Diffuse $\gamma$-rays, produced by cosmic rays interacting with matter in the Galaxy plane, are expected to contribute to the ARGO-YBJ result.
According to the measurement of diffuse $\gamma$-ray flux from the inner Galactic plane using ARGO-YBJ data \citep{ma11}, this contribution
to the flux from HESS J1841$-$055 in the five intervals is less than 4\%.
Assuming the HESS shape for the source instead of the symmetrical two-dimensional Gaussian shape, the flux would only vary of 2.2\%. HESS J1837$-$069 is the nearest VHE $\gamma$-ray source with an angular distance of 1.62 degrees. The flux from HESS J1837$-$069 at energies above 1 TeV is 17\% that of the Crab with spectral index $-$2.27 \citep{aharon06b}. Its contributions to the five intervals are estimated to be 5.8\%, 2.7\%, 1.0\%, 0.7\%, and 0.2\%, respectively.
The second nearest source is HESS J1843-033, whose flux is still unknown.
A hot spot with a marginal significance of 4.1 s.d. is
observed near its position \citep{hoppe07}.
With an angular separation of 2.6 degrees, its contribution is estimated to be
lower compared with HESS J1837$-$069.
The contribution from other known VHE $\gamma$-ray sources is negligible. Moreover, an estimate of the systematic error of ARGO-YBJ is described in \citep{barto12b}. With an incomplete list of possible causes, such as time resolution variation, event rate variation with environment parameters, and pointing error, the systematic error for point sources is found to be less than 30\%, and is lower for extended sources. Thus, the systematic error of ARGO-YBJ alone is not enough to explain the discrepancy.

HESS J1841-055 is observed by ARGO-YBJ only at high zenith angles ($\theta >35.7^{\circ} $), while the observation of the Crab is possible also at low zenith angles. This difference may cause some systematic errors when estimating the spectrum of HESS J1841-055. To check a possible systematic error, observations of the Crab at zenith angles higher than 30$^{\circ}$ are used. With this selection, the average zenith angle is about the same as that of HESS J1841-055.
The result is that the Crab spectral index varies from (2.58$\pm$0.07) to
(2.52$\pm$0.21), and the flux above 1 TeV is (35$\pm$28)\% higher.
Due to the large statistical error, we cannot exclude a systematic effect causing the difference of flux. However, even taking this systematic error into account, the flux of HESS J1841-055 observed by ARGO-YBJ is still about twice that determined by HESS.

On the other hand, the discrepancy is similar to that found for the two extended sources MGRO J1908+06 and MGRO J2031+41 \citep{barto12c, barto12b}. The fluxes measured by the EAS arrays Milagro and ARGO-YBJ are much higher than that determined by the Cherenkov arrays. Since a good agreement has been achieved on the ``standard candle" Crab Nebula, some systematic differences between the two techniques should exist only for extended sources. As pointed out by \cite{abdo12}, due to their limited FOV, Cherenkov telescopes might count the extended emission as background, especially when using the ``wobble mode'' to estimate this latter. It is worth noting that the ``wobble mode'' was used when HESS observed HESS J1841$-$055, and the source was offset by 0.7$^{\circ}$ \citep{aharon08}. The ``reflected-region technique'' is used to estimate the background for spectra in a region partially overlapping the extended source. As a result, HESS would measure an emission fainter than that measured by ARGO-YBJ.

Different scenarios have been proposed to explain the emission mechanism
of TeV photons. VHE $\gamma$-rays can be produced via inverse-Compton of
background photon fields by high energy electrons, or, in hadronic
models, by inelastic proton-proton or proton-photon interactions. In
both scenarios, X-ray and radio synchrotron emissions are
expected, therefore the lack of a low energy counterpart for HESS J1841$-$055
poses the question about the nature of the emission mechanism.
\cite{aharon08} searched for counterparts responsible of the VHE
$\gamma$-ray emission and discussed the possible association with six
candidates, marked in Figure 4, which is a zoom of Figure 1 around
HESS J1841$-$055.
Three of them are the pulsars PSR J1838$-$0549, PSR J1841$-$0524
and PSR J1837$-$0604, of which only the last has a high enough spin-down
flux ($\dot E /D^2 = 5.2\times 10^{34}$ erg s$^{-1}$ kpc$^{-2}$) to be a
counterpart candidate. This source is at the boundary of the HESS region, but
not far from the center of gravity of the ARGO-YBJ signal.
However, as pointed out in \citep{aharon08}, since the TeV emission is usually
attributed to a relic population of electrons, some contribution can be
expected also from the other pulsars if they had a much higher spin-down
luminosity in the past. No catalogued PWNs at longer wavelengths are
associated to these pulsars, however, according to the recent calculations
of \citep{tibolla12}, during their evolution ancient PWNs ($>>$10 kyr) might
appear as GeV-TeV $\gamma$-ray sources without X-ray counterpart.
The three other catalogued objects located inside the HESS uncertainty
region are the SNR G027.4 (also known as Kes73), the high-mass XB AX
J1841.0$-$0536 and the diffuse source G26.6-0.1. The SNR Kes73 lies at the
edge of the TeV emission region. The point-like nature of AX J1841.0$-$0536,
the only soft $\gamma$-ray source detected within the HESS J1841$-$055 error
ellipse, its variability and the required luminosity (about $10^{36}$ erg
s$^{-1}$ according to the ARGO-YBJ data and assuming a distance of 6.9 kpc as
inferred in \citep{sguer09}) exclude its association with the entire
emission from the extended HESS source. Also the diffuse source
G26.6-01, at only 1.3 kpc and well inside the emission region, could be
responsible at least for part of the TeV flux.
In Figure 4 are also reported four GeV $\gamma$-ray sources from the
$Fermi$-LAT second source catalogue within the extension of HESS J1841$-$055:
2FGL J1839.3$-$0558c, 2FGL J1836.8$-$0623c, and the two diffuse sources
2FGL J1839.0$-$0539 and 2FGL J1841.2$-$0459c \citep{abdo12a}. Moreover, the two
event clusters at energies above 100 GeV found by \citep{neron10} are shown.
Three GeV sources are within the two event clusters, suggesting that they may
be also VHE emitters: 2FGL J1841.2$-$0459c is coincident with the SNR Kes 73,
while 2FGL J1839.3$-$0558c and 2FGL J1839.0$-$0539 are spatially associated
to the PSR J1838-0549 and the diffuse X-ray G26.6$-$0.1, respectively.
As remarked in \citep{tibolla12}, the recent observation by $Fermi$-LAT of GeV
sources not firmly associated to X-ray counterparts
suggests that VHE unidentified sources can be explained as ancient PWNs.

An hadronic scenario is proposed in \citep{neron12}. These
authors consider the extended $\gamma$-ray emission produced by high-energy
cosmic rays escaping from the source and diffusing in the interstellar
medium (ISM). The $\gamma$-ray emission should result from the interaction of
these cosmic rays with the ISM particles. Such extended emission regions
should be visible as VHE $\gamma$-ray sources with fluxes of order
$10^{-11}$ erg cm$^{-2}$ s$^{-1}$ above 100 GeV. From the analysis of the
$Fermi$-LAT data they suggest the young nearby pulsar PSR J1841$-$0524 as a
possible low energy counterpart of HESS J1841$-$055. However, as already
stated, due to the energy balance, this association is not without problems.
The proton/nuclei contribution to the extended $\gamma$-ray flux should
generate a comparable flux of TeV neutrinos with a spectrum expected to follow
the $\gamma$-ray spectrum. Thus, the observation of high energy neutrinos from
the HESS source could provide a crucial test to this model. A search for
individual neutrino sources over a large fraction of both the northern and
southern skies has been carried out by the IceCube detector in the 40-string
configuration \citep{abbasi11}. The large background from atmospheric
muons reduces the IceCube sensitivity to neutrino sources in the southern
sky at TeV energies, thus the derived upper limits are not stringent enough
to constrain the hadronic scenario.
Data from the combined operation of IceCube and AMANDA have been used to
scan for sources in the Galactic plane \citep{abbasi12} with a
neutrino flux sensitivity of about $10^{-11}-10^{-12}$ erg cm$^{-2}$ s$^{-1}$
at TeV energies. However, the surveyed range of Galacic longitude
($36^{\circ} < l < 210^{\circ}$) does not include the region where
HESS J1841$-$055 is located.

In the case of hadronic scenarios one expects the source extension to be much larger than seen by Cherenkov telescopes (up to the degree scale). Therefore, the
lower angular resolution and the large FOV of ARGO-YBJ allows the collection of photons from a larger source area. This could partially explain the discrepancy in flux with the HESS Cherenkov telescope.

Recently, \cite{giaci12} found that diffusion of cosmic rays and electrons around point sources is strongly anisotropic and shows filamentary structures, which may cause the shift of the centroid position between HESS and ARGO-YBJ.

\section{Conclusions}

Since November 2007 the ARGO-YBJ experiment is monitoring with
high duty cycle the northern sky at TeV photon energies.
Using data up to July 2012, an excess with statistical significance of 5.3 s.d.
is detected from the direction of the unidentified source HESS J1841$-$055.
The source location and extension are
consistent with those determined by HESS, however the measured flux above
1 TeV is about 3 times higher.
This discrepancy, already found in the observation of other extended sources,
could origin from the different techniques used in the
background estimation for extended sources with ARGO-YBJ and HESS data.
The extended morphology of HESS J1841$-$055 and the presence
of several sources within the 90\% confidence error region suggests
contributions from more than one of them, but so far no clear counterparts in
lower-energy wavebands can be identified.
However, the possibility of a GeV-TeV $\gamma$-ray source without any
counterpart can not be excluded.
Both leptonic and hadronic productions of $\gamma$-rays have
been proposed, but it is not easy to distinguish between the two
contributions basing only on the $\gamma$-ray data.
The current upper limits to the neutrino flux from the HESS J1841$-$055
region are too high to test the hadronic model.
Further multiwavelength observations from radio to GeV energies and data
from neutrino telescopes of suitable sensitivity are needed in order to
disentangle between the different emission possibilities.

\acknowledgments
 This work is supported in China by NSFC (No.10120130794, No.11205165),
the Chinese Ministry of Science and Technology, the
Chinese Academy of Sciences, the Key Laboratory of Particle
Astrophysics, CAS, and in Italy by the Istituto Nazionale di Fisica
Nucleare (INFN).

We also acknowledge the essential supports of W.Y. Chen, G. Yang,
X.F. Yuan, C.Y. Zhao, R. Assiro, B. Biondo, S. Bricola, F. Budano,
A. Corvaglia, B. D'Aquino, R. Esposito, A. Innocente, A. Mangano,
E. Pastori, C. Pinto, E. Reali, F. Taurino and A. Zerbini, in the
installation, debugging and maintenance of the detector.


\clearpage
\begin{figure}
\plotone{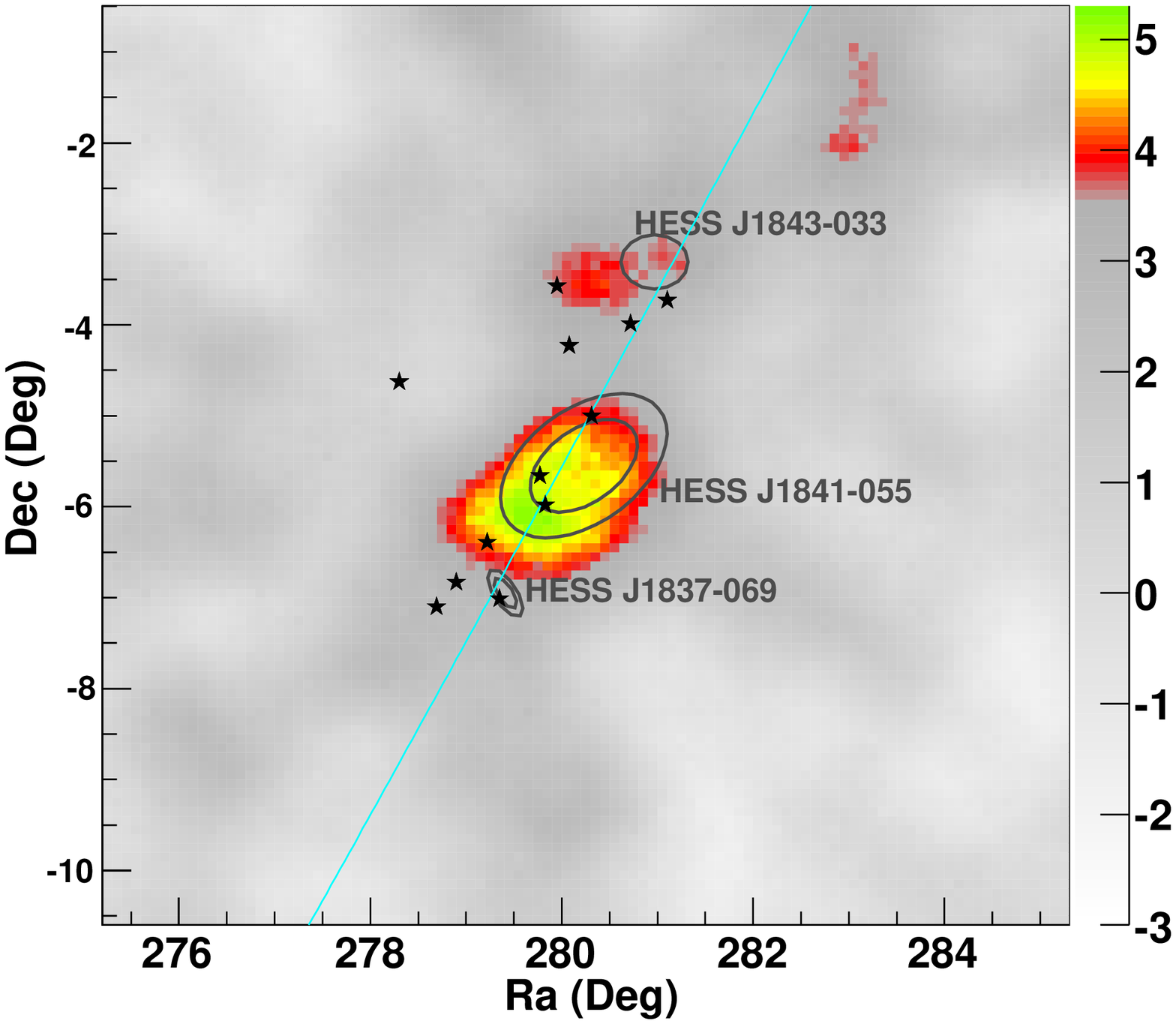}
\caption{
The significance map around HESS J1841$-$055 as observed by the ARGO-YBJ
experiment. The two ellipses for HESS J1841$-$055 and HESS J1837$-$069
indicate their positions and the 68\% and 90\% contours of their extension
regions \citep{aharon08}. The position and  possible extension of HESS J1843-33 are  marked with ellipse \citep{hoppe07}.
The  stars mark the location of the GeV $\gamma$-ray sources around HESS J1841$-$055 in the second $Fermi$-LAT catalog \citep{abdo12a}. The solid line indicates the Galactic plane. }
\label{fig1}
\end{figure}

\begin{figure}
\plotone{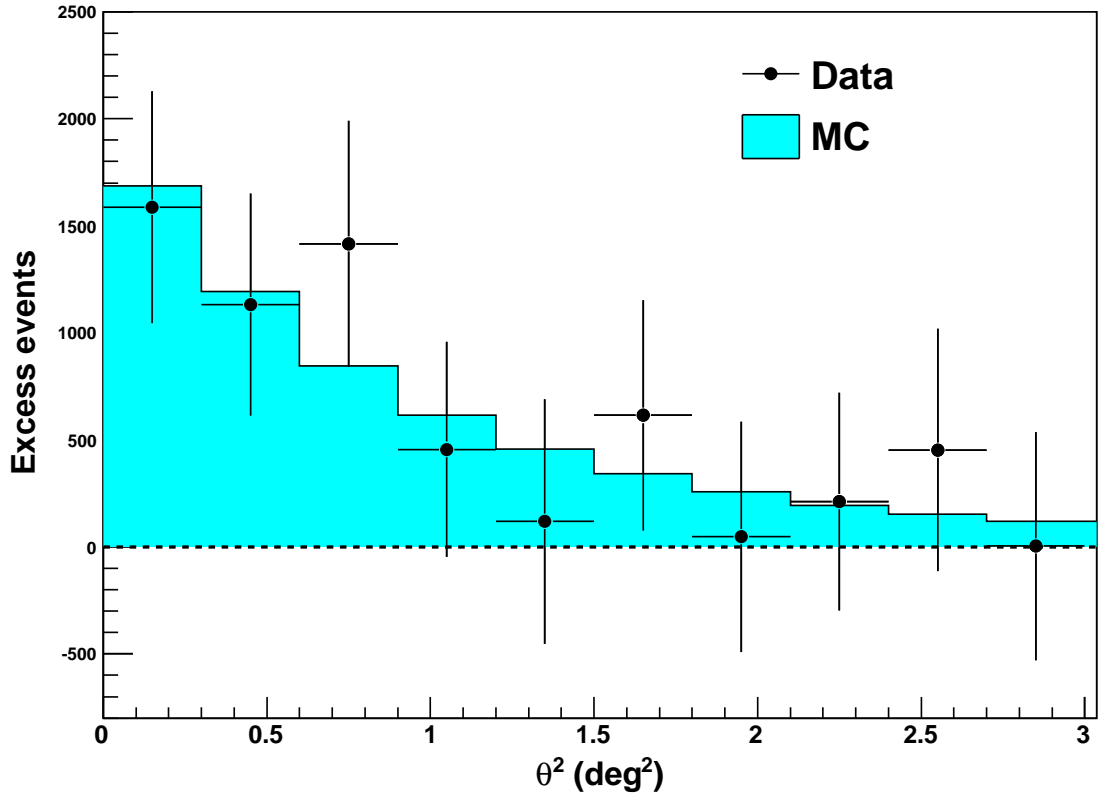}
\caption{Distribution of $\theta^{2}$ for the number of excess events around
HESS J1841$-$055. The filled region outline the best fit to simulated data
assuming a symmetrical two-dimensional Gaussian shape with
$\sigma=0.40^{\circ}$.}
\label{fig2}
\end{figure}

\begin{figure}
\plotone{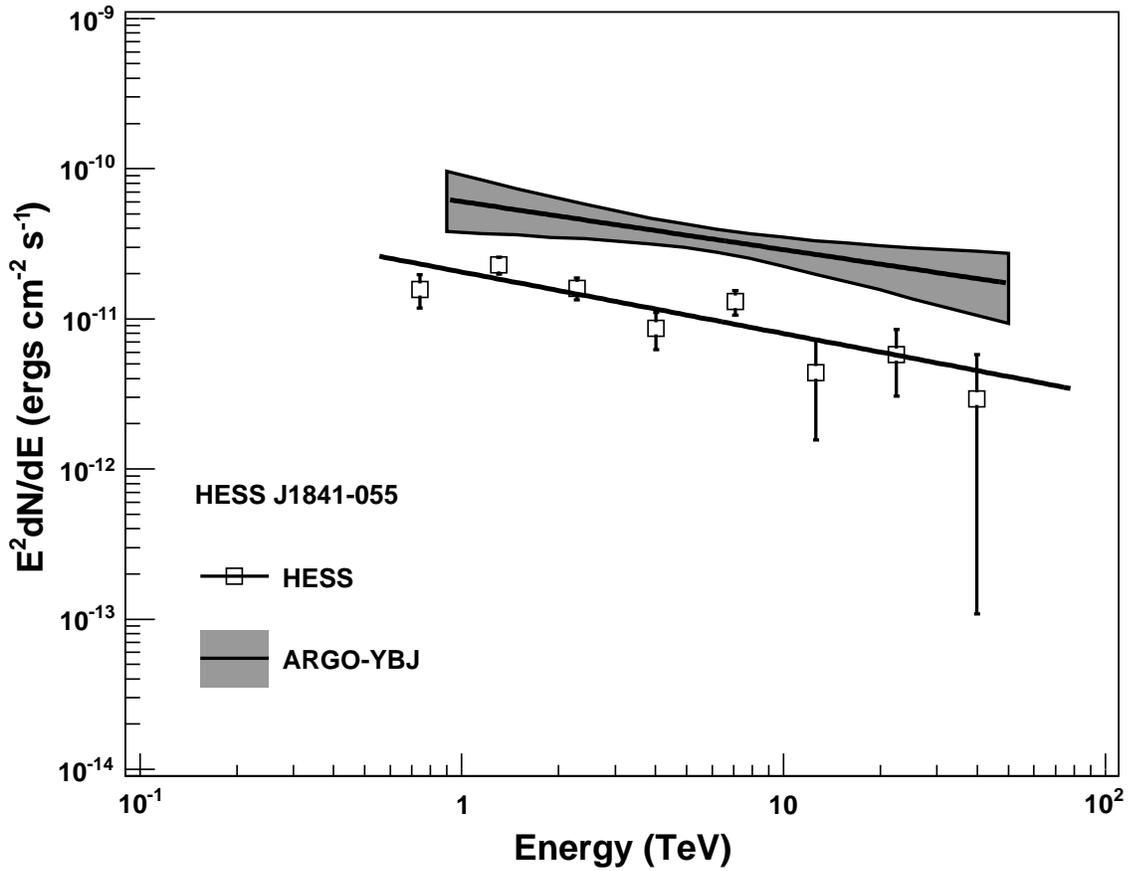}
\caption{Energy density spectrum of HESS J1841$-$055 as measured by the
ARGO-YBJ experiment: the solid line and shaded area indicate the differential
energy spectrum and the 1 s.d. error region. The spectum
measured by HESS \citep{aharon08} is also reported for comparison.
Only statistical errors are shown.}
\label{fig3}
\end{figure}

\begin{figure}
\plotone{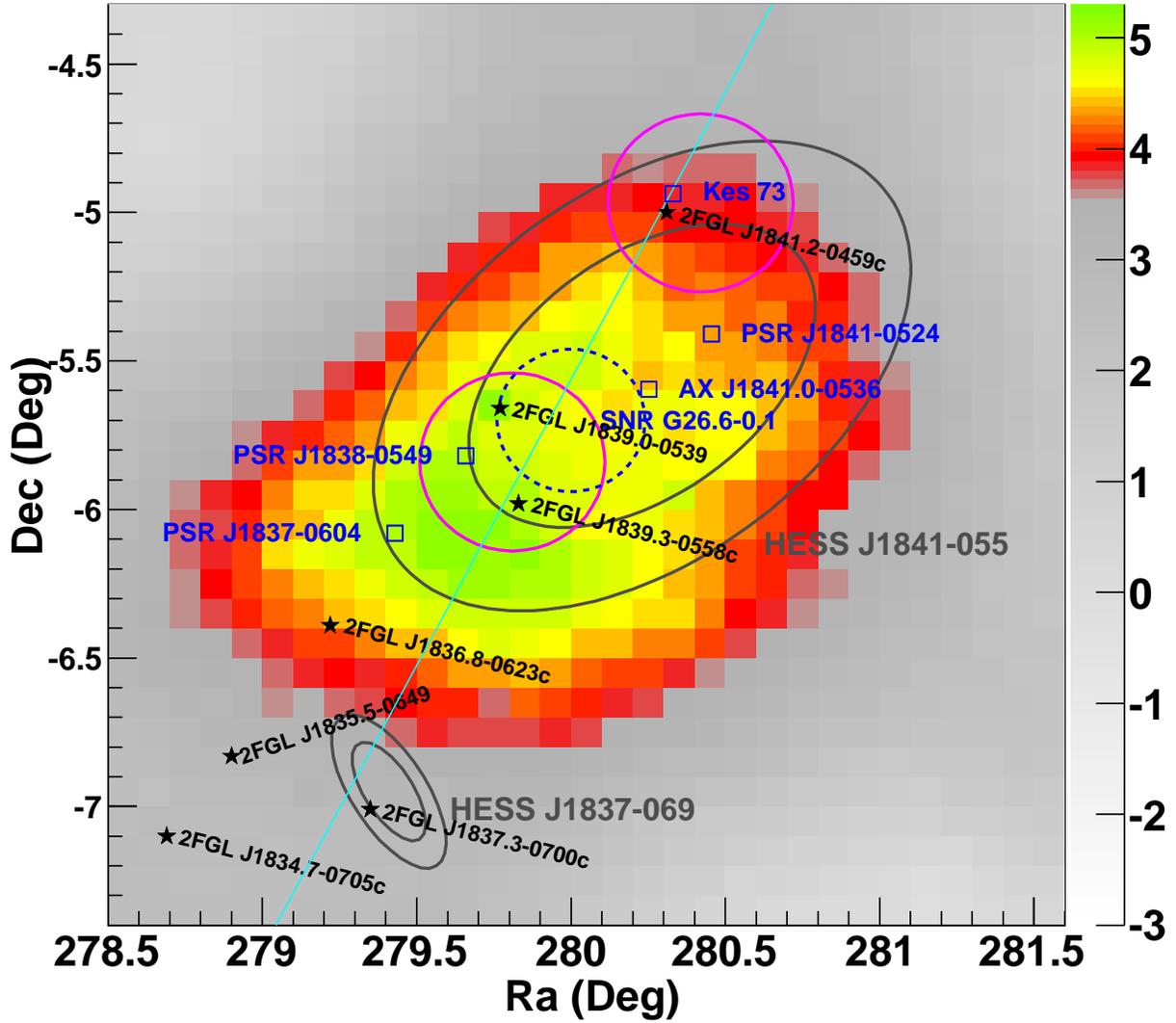}
\caption{Zoom of Figure 1 around HESS J1841$-$055. The squares and the dashed circle indicate the position of the candidates reported in \citep{aharon08}. The circles indicate the two event clusters found in \citep{neron10} at energies above 100 GeV. The ellipses and stars are the same as in Figure 1. The solid line indicates the Galactic plane. }
\label{fig3}
\end{figure}

\clearpage

\end{document}